\documentclass[aps,prl,twocolumn,groupedaddress]{revtex4}
\usepackage{graphicx}% Include figure files
\usepackage{dcolumn}% Align table columns on decimal point
\usepackage{bm}% boldmath \citestyle{prb}

\begin{document}

\title{A new microscopic mechanism for secondary relaxation in glasses}

\author{L.C. Pardo$^1$, M. Rovira-Esteva$^1$, S. Busch$^2$, M.D. Ruiz-Martin$^1$, J.Ll. Tamarit$^1$, T. Unruh$^2$}

\affiliation{$^1$Grup de Caracteritzaci\'{o} de Materials,
Departament de F\'{\i}sica i Enginyieria Nuclear, ETSEIB,
Universitat Polit\'ecnica de Catalunya, Diagonal 647, 08028
Barcelona, Catalonia,Spain}
\affiliation{$^2$ Physik Department E13 and
Forschungsneutronenquelle Heinz Maier-Leibnitz (FRMII), Technische
Universit\"at M\"unchen, Lichtenbergstr.\ 1, 85748 Garching,
Germany}

\title{Bayesian Analysis of QENS data:
From parameter determination to model selection}

\begin{abstract}

The extraction of any physical information from
quasielastic neutron scattering spectra is generally done by
fitting a model to the data by means of $\chi^2$ minimization
procedure. However, as pointed out by the pioneering work of D.S.
Sivia et al. \cite{Sivia_QENS}, also another probabilistic approach
based on Bayes theorem \cite{Sivia_book} can be employed. In a
nutshell, the main difference between the classical $\chi^2$
minimization and the Bayesian approach is the way of expressing
the final results: In the first case, the result is a set of
values of parameters with a symmetric error ($P_i \pm
\varepsilon_i$) and a figure of merit such as $\chi^2$, whereas in
the second case the results are presented as probability distribution
functions (PDF) of both, parameters and merit figure. In this
contribution, we demonstrate how final PDFs are obtained by
exploring all possible combinations of parameters that are
compatible with the experimental error. This is achieved by
allowing the fitting procedure to wander in the parameter space
with a probability of visiting a certain point $P=exp(-\chi^2/2)$,
the so called Gibbs sampling. Three advantages of this method will
be emphasized: First, correlations between parameters are
automatically taken into account, which implies, for example, that
parameter errors are correctly calculated, correlations show up in
a natural way and ill defined parameters (i.e. parameters for
which data only support the calculation of lower or upper bounds)
are immediately recognized from their PDF. Second, it is
possible to calculate the likelihood of a determined physical
model, and therefore to select the one among many that fits the
data best with a minimal number of parameters, in a correctly
defined probabilistic way. Finally, in the case of a low count
rate where the Gaussian approximation to the Poisson statistics
fails, this method can also be used by simply redefining
$\chi^2$.

\end{abstract}

\maketitle

% Hier bitte den Inhalt des Artikels einfügen.
% Please insert the content of your submission here.

\section{Introduction}

Science is based on the success of an hypothesis
to describe experimental results, i.e. is based on the amount of
"truth" and "falsity" of an hypothesis when contrasted with
experimental results \cite{popper}. In order to find a
quantitative method to determine this "amount of truth",
hypotheses in science should at the end be reduced to a
mathematical expression depending on a set of parameters with some
physical meaning. The "amount of truth" is then determined by
fitting the mathematical model to some experimental data. The
general method to do so is to minimize the squared distance
between experimental data and the points generated by the
mathematical model. Furthermore, taking also into account the
error associated with experimental data, a figure of
merit $\chi^2$ can be defined
\begin{equation}
\label{eqchisquare}
\chi^2=\sum^{n}_{k=1}\frac{(H_k\{P_i\}-D_k)^2}{\sigma_k^2}
\end{equation}
where $n$ is the number of experimental points and $m$ is the number of parameters, $D_k$ ($k$=1,... $n$)
are the experimental data, $H_k\{P_i\}$ ($k$=1,... $n$) are the values
obtained from our hypothesis (the mathematical model) using the
$\{P_i\}$ ($i$=1,... $m$) set of parameters contained in the model,
and $\sigma_k$ ($k$=1,... $n$) are the experimental errors associated with the respective $D_k$.
Therefore, the fitting procedure has a twofold
goal: first, to find the set of parameters $\{P_i\}$ which describes the experimental data best, and second, using this set of
parameters, to define a figure of merit which quantifies the "amount
of truth" of the proposed hypothesis. In order to be able to
compare different hypotheses with different numbers of parameters
it is reasonable to define a figure of merit which penalizes additional parameters such as the \emph{reduced $\chi^2$} defined as:
\begin{equation}
\label{eqredchisquare} \chi_{\nu}^2=\frac{\chi^2}{n-m}
\end{equation}
where $n$ is the number of experimental points and $m$ is the number
of parameters, so $n-m$ is the number of degrees of
freedom. The aforementioned way to quantify how good experimental data are described by a hypothesis is based on what is called a
"frequentist" approximation to the problem \cite{frequentist}, and
has many drawbacks  associated with both the fitting procedure and
the way to quantify the correctness of the hypothesis  describing
experimental data.

Data fitting is usually done by minimizing  the aforementioned
$\chi^2$ (equation \ref{eqchisquare}) using the Levenberg-Marquardt
algorithm, which aims to find the minimum of the $\chi^2\{P_i\}$
hypersurface. Unfortunately, usually local minima make the algorithm
unable to find the absolute minimum. For this reason, this
method can find a final solution only when the algorithm is initialized with parameters near
the global minimum. The final solution is then characterized by a set of parameters with an
associated error ($P_k\pm\varepsilon_k$) and the figure of merit $\chi_{\nu}^2$. This way of quantifying the best fit to the data
is based on the supposition that there is only one minimum in the
$\chi^2(P_k)$ hypersurface compatilbe with data error, and that the functional dependence of
$\chi^2(P_k)$ is quadratic on each parameter (i. e. one can
stop at the second term of a Taylor expansion of the obtained
minimum), and thus allowing only symmetric errors. Moreover, errors
are usually calculated disregarding possible
correlations between them \cite{diagonal} and are thus generally underestimated.

We present in this work a method both to perform fittings and to
analyze results based exclusively on probability by using what is
called Bayesian inference. The main difference with the previously
exposed frequentist method is the absence of \emph{any}
supposition on the $\chi^2\{P_i\}$ landscape which will rather be
explored using the probabilities determined from  experimental
data. The method results in a
different way to express fitted parameters and the figure of merit
showing all the complexity of the final solution: they become
Probability Distribution Functions (PDFs) obtained directly from
exploring the $\chi^2\{P_k\}$ hypersurface.

The paper will be
organized as follows: first the ubiquitous $\chi^2$ will be
defined using exclusively probability theory, and on this basis a
method to sample the $\chi^2\{P_k\}$ hypersurface will be
presented: the Gibbs sampling. We will then refer on how both the frequentist and Bayesian methods select an hypotheses among others, stressing the advantages of using the second approach. Finally the presented method implemented in the FABADA package \cite{fabada} will be applied to three real cases related to neutron scattering
each stressing  different aspects of the proposed method. In the first example, the importance of letting parameters free or fixed in the fitting process will be stressed. The second example will focus on the PDF obtained from a set of data fitted simultaneously, and model selection will be addressed in the third example.

\section{What is behind the ubiquitous $\chi^2$?}

The objective of the so called \emph{Bayesian methods} \cite{Sivia_QENS,Sivia_book} is to find the
probability that a hypothesis is true given some experimental
\emph{evidence}. This is done taking into account both our
\emph{prior} state of knowledge concerning the hypothesis, and the
\emph{likelihood} that the data is described by the proposed hypothesis. Using probability notation, and only considering
the case that the experiment consists of a series of data $D_k$ and
that the hypothesis is represented by $H_k$, we can relate the aforementioned probabilities using the Bayes theorem \cite{bayes}:

\begin{equation}
\label{eqbayes} P(H_k \mid D_k)=\frac{P(D_k \mid
H_k)P(H_k)}{P(D_k)}
\end{equation}

where $P(H_k \mid D_k)$ is called the  \emph{posterior}, the
probability that the hypothesis is in fact describing the data.
$P(D_k \mid H_k)$ is named the \emph{likelihood}, the probability
that our data is well described by our hypothesis. $P(H_k)$ is
called the \emph{prior}, the knowledge we have beforehand about the
hypothesis, and $P(D_k)$ is a normalization factor to assure
that the integrated posterior probability is unity. In the
method here presented we will assume no prior knowledge (maximum
ignorance prior \cite{Sivia_book}), and in this special case Bayes
theorem takes the simple form:

\begin{equation}
\label{eqlikeli}
P(H_k \mid D_k) \propto P(D_k \mid H_k)\equiv L
\end{equation}
where $L$ is a short notation for likelihood.

We need first to find the likelihood that one data point $D_k$ is
described by the mathematically modeled hypothesis $H_k$. In a counting
experiment such as those related to neutron scattering this
probability follows a Poisson distribution

\begin{equation}
\label{eqlikelip}
P(D_k \mid H_k)=\frac{{H_k}^{D_k} \; e^{-H_k}}{D_k!}.
\end{equation}

\begin{figure}
\includegraphics[width=\columnwidth]{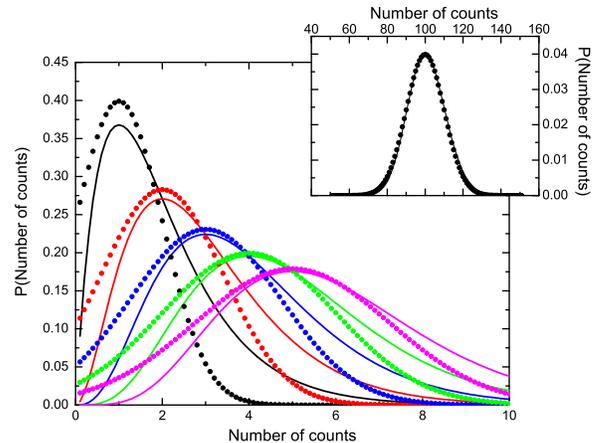}
\caption{Poisson statistics followed by a counting experiment such
as a neutron scattering experiment (lines). For an increasing number
of counts, the Poisson distribution can be approximated by a Gaussian
function (points) with $\sigma = \sqrt{n}$, being $n$ the number of
counts.}
\label{pvg}
\end{figure}

Nevertheless, for a high enough number of counts, the Poisson PDF
can be well approximated by a Gaussian one with $\sigma=\sqrt{D_k}$ as
it is shown in figure \ref{pvg}, and hence the likelihood that
the set of data points $D_k$ is correctly described by the
hypothesis $H_k$ can be written as

 \begin{eqnarray*}
\label{eqlikeli2}
L &\propto& \prod_{k=1}^{n}\exp\left[-\frac{1}{2}\left( \frac{H_k-D_k}{\sigma_k}\right)^2\right] \\ &=& \exp\left[{-\frac{1}{2}\sum_{k=1}^{n} \left( \frac{H_k-D_k}{\sigma_k}\right)^2}\right]=\exp\left(-\frac{\chi^2}{2}\right).
\end{eqnarray*}

Therefore, we have found the meaning of
the ubiquitous $\chi^2$ based only on probabilistic grounds: it is related to the probability that a
certain set of data is well described by an hypothesis, and
hence the goal of minimizing $\chi^2$ is finding a set of
parameters that maximizes the likelihood
 associated with the
proposed mathematical model. The probability theory behind $\chi^2$ allows therefore also to deal with the case of experiments with only few counts where the Gaussian
approximation is not valid anymore and the Poisson distribution
must be employed simply by redefining $\chi^2$ as
\begin{equation}
\label{eqchipoi}
\chi^2= -2 \cdot \sum_{k=1}^n \ln \left[ \frac{{H_k}^{D_k}e^{-H_k}}{D_k!} \right]
\end{equation}

\section{The Bayesian method: Gibbs sampling of parameter space}

The probabilistic understanding
of $\chi^2$ makes it possible to define a unique method, first to fit
the experimental data, and then to analyze the obtained results,
using a Markov Chain Monte Carlo (MCMC) technique. A
set of parameters $P_i^{new}$ is generated from an old set $P_i^{old}$ by randomly changing one of the parameters \cite{jumpchange}. The probability to accept the new set of parameters is given by
\begin{equation}
\label{eqjump}
\frac{P(H(P_i^{new})\mid D_k)}{P(H(P_i^{old})\mid D_k) }=\exp\left(-\frac{\chi^2_{new}-\chi^2_{old}}{2}\right)
\end{equation}

where $\chi^2_{new}$ and $\chi^2_{old}$ correspond to the  $\chi^2$ (as defined in equation \ref{eqchisquare}) for
the new and old set of parameters. This way of exploring the
parameter space (called Gibbs sampling) is similar to the way used to
find the possible molecular configurations of a determined system at a given temperature
using the classical Monte Carlo method: the values of physical
constants such as the potential energy will  in fact be a PDF
related to all the configurations explored by the Monte Carlo
method. It is therefore possible to relate energy to $\sum(H_k-D_k)^2$, the magnitude giving
information about the fit quality of the hypothesis with respect to the data,
and temperature to the error associated with
the data ($T\sim\sigma^2$). This wat of exploring the parameter space has two main advantages:

\begin{itemize}
\item In the fitting process, the Bayesian method is able to accept a new set of parameters that do not decrease $\chi^2$, if this change is compatible with the experimental error and therefore does not get stuck in local minima as the Levenberg-Marquardt algorithm. In other words, the presented method is able to go "uphill" in the $\chi^2\{P_i\}$ hypersurface if the barrier is compatible with the error. Nevertheless, in order to
avoid the presented algorithm to get stuck even in the case when barriers are greater than those associated to experimental error a
simulated annealing can be used. This algorithm calculates a
fictitious $\chi^2=\sum_{k=1}^{n}\frac{(H_k-D_k)^2}{T \sigma^2}$
where T is a constant defined to artificially increase the
experimental error, and by similitude with classical Montecarlo
simulation is named as "temperature". Fittings are then
started at high temperature, and the system is relaxed by lowering
the temperature up to T=1.
\item Concerning the analysis of the results obtained by the fitting, the exploration of the whole parameter space compatible with data using the MCMC method allows both to find the PDF associated with the likelihood directly related the figure of merit $\chi^2$ (see equation \ref{eqlikeli}), and the parameters, taking into account possible correlations between them, or minima not describable by a quadratic approximation.
\end{itemize}

\section{Model selection}
\begin{figure}
\includegraphics[width=\columnwidth]{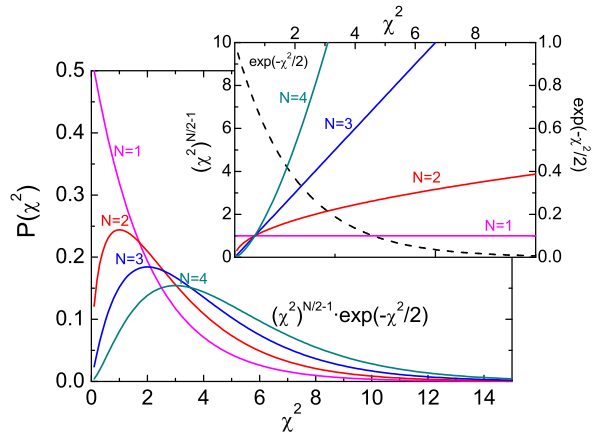}
\caption{$\chi^2$ distribution ($(\chi^2)^{N/2-1}\exp(-\chi^2/2)$) for an increasing number of fitting parameters N. The inset shows the terms associated to the quality of the fit $\exp(-\chi^2/2)$ together with the one depending on the number of parameters of the model $(\chi^2)^{N/2-1}$.}
\label{chisqdist}
\end{figure}

Data can usually be described by more than one hypothesis, each implying a different physical mechanism to explain experimental results. Albeit the importance to perform model selection accurately, vague arguments are usually given to prefer a model among others and usually no quantitative arguments are given to justify why an hypothesis is preferred, although it is possible to do so using both the frequentist and Bayesian methods.
Model selection can be performed using the frequentist approach by using the $\chi_{\nu}^2$ figure of merit
(see equation \ref{eqredchisquare}) which takes into account the addition of parameters to a model by dividing $\chi^2$ by the degrees of freedom. Therefore, if two models fit the data with equal success, i.e. with the same $\chi^2$, the model with less parameters (with the smallest $\chi_{\nu}^2$) will be favored. In some sense this is nothing but quantifying the Ockham's razor principle: it is necessary to shave away unnecessary assumptions (parameters). Model selection performed by using $\chi_{\nu}^2$ has the same drawbacks as the determination of parameter errors: we suppose that there is a single minimum in $\chi^2\{P_i\}$, that this  minimum parabolic depends on all parameters and that there are no correlations between parameters. In fact, if these three suppositions are accomplished, then the PDF of the $\chi^2$ reads \cite{Sivia_book}
\begin{equation}
\label{eqchidist}
P(\chi^2) \propto (\chi^2)^{N/2-1}\exp(-\chi^2/2)
\end{equation}
 $N$ is, in this simple case, the number of parameters. In figure \ref{chisqdist} the chi-square distribution for increasing degrees of freedom (number of parameters) is shown. As can be seen in the inset from figure \ref{chisqdist}, this distribution has a term which is independent from the number of parameters, $\exp(-\chi^2/2)$, and that decreases together with the quality of the fitting, or when the error associated with the experimental data $\sigma_k$ increases. The term $(\chi^2)^{N/2-1}$, depending on the number of degrees of freedom, increases exponentially with the number of parameters, displacing the maximum of the $\chi^2$ distribution to higher values. Therefore, even using the frequentist approach, the aforementioned preference for models that fit equally well the data with the minimum number of parameters is based on probability theory: those models with the maximum in the $\chi^2$ distribution placed at lower values will be preferred. The Bayesian method finds in a natural way the PDF of $\chi^2$ by exploring the parameter space without the suppositions made in the frequentist approximation, hence the obtained PDF will in general not follow the $\chi^2$ distribution described by equation \ref{eqchidist}.

\section{Examples}

\subsection{Determining the intramolecular structure of CCl$_4$}

Molecular structure can be calculated from diffraction experiments
by fitting the high $q$-range of the scattering function $S(q)$ to the
following equation (see \cite{liquids}):

\begin{equation}
\label{eqsq}
S(q)=h \sum_{i,j}^{m} b_i^{coh} b_j^{coh} \frac{sin(qr_{ij})}{qr_{ij}}\exp{\frac{\left(- l_{ij}^2 q^2\right)}{2}}
\end{equation}

where $b_i^{coh}$ are the coherent cross sections for each element,
$r_{ij}$ are the intramolecular distances and $l^2=\left<u_{ij}^2\right>$ are the vibrational
Mean Square Displacements (MSD) between elements $i$ and $j$, and $h$ is a scaling factor.

In the
proposed example, our objective is to calculate the intramolecular
structure of carbon tetrachloride (one of the first molecular
liquids studied by diffraction methods). The determination
of its molecular structure implies to obtain the distance between
carbon and chlorine atoms, the Cl-Cl distance is fixed by the
tetrahedral symmetry, and the MSD between chlorine atoms and
carbon and chlorine atoms.

\begin{figure}
\includegraphics[width=\columnwidth]{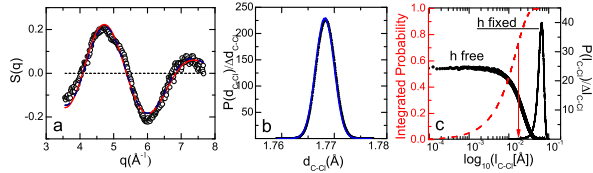}
\caption{(a) High momentum transfer scattering function for CCl$_\mathrm{4}$, where the main contribution to S(q) is that related to the molecular structure.
Lines are the best fits to S(q) setting the scale factor fixed and free in the fitting procedure. (b) Probability
distribution functions obtained for the distances between carbon
and chlorine atoms $r_\mathrm{CCl}$, and Mean Square Displacement between carbon
and chlorine atom ($l_\mathrm{CCl}$) for both cases (fixed $h$, points, and free $h$, line). $PDF(l_\mathrm{CCl})$ has been represented in logarithmic scale to show the different length scales explored by the Bayesian method. In addition, for $l_\mathrm{CCl}$ using a $h$ free fitting the integrated probability is shown, arrow points an integrated probability of 0.68 following the standard definition of errors in the frequentist approximation \cite{errors}.}
\label{gccl4}
\end{figure}

Experiments were performed at  the diffractometer D1b in the Institute Laue Langevin (Grenoble, France) using a
wavelength of $\lambda=2.52\mathring{A}$ (see \cite{CCl4}). Figure
\ref{gccl4}a shows a good agreement between experimental data and
two fittings of equation \ref{eqsq}, one with a fixed scale factor $h$, and the other with $h$ as a free parameter. Figures \ref{gccl4}b,c show
the PDF from parameters $r_\mathrm{CCl}$ and $l_\mathrm{CCl}$ obtained through the two aforementioned fittings. Concerning the
$r_\mathrm{CCl}$ PDF, we can immediately see that its determination is
robust since both fixed and free scale factor $h$ lead to the same PDF. On the contrary, the PDF associated with
$l_\mathrm {CCl}$ is sensible
to the way we have performed the fitting: if the scale factor is
fixed we obtain a most probable value for this parameter
($l_{CCl}=0.066 \mathring{A}$), but for a free scale factor $h$ only a maximum
value for $l_\mathrm{CCl}$ can be obtained due to the correlation between both parameters ($h$ and $l_\mathrm{CCl}$). Defining the upper limit as that for which the integrated probability is 0.682 (as errors are usually defined in the frequentist approach \cite{errors}) the upper limit $l_\mathrm{CCl}=0.02 \mathring{A}$ can be determined from the cumulative distribution function (see fig. \ref{gccl4}).

This example shows the main difference compared to the frequentist approximation: the results are presented as PDF. This has the
advantage that, as it happens with the determination of $l_\mathrm{CCl}$
leaving $h$ free, the result to our parameter determination can be
expressed as a limit for the parameter, which is impossible with
the frequentist approximation.

\subsection{Parameter estimation: isotropic rotation}
\begin{figure}
\includegraphics[width=\columnwidth]{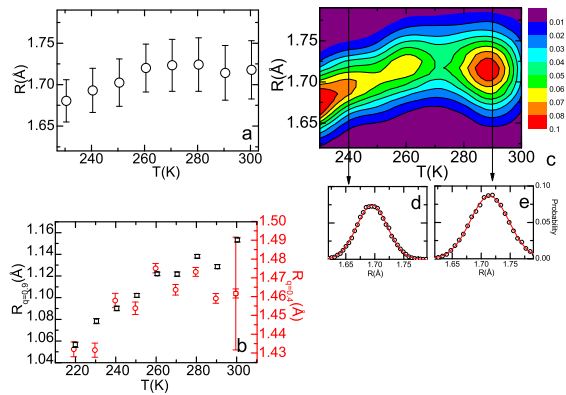}
\caption{Radius of rotation obtained (a) from the Bayesian fitting procedure applied to all $S(q,\omega)$ spectra and (b) from the frequentist fitting to each $q$-value individually (here only $q=0.4 \mathring{A}$ and $q=0.9 \mathring{A}$ values are shown). An error bar obtained from the Bayesian method is plotted in (b) for $S(q=0.4,\omega)$ at $T=300K$. In (c) the PDF for the radius of rotation is also shown as a function of the temperature ($P(R,T)$) with two cuts ($P(R)$) for $T=240 K$ and $T=290 K$.}
\label{Rgirobayesg}
\end{figure}

Quasi Elastic Neutron Scattering (QENS) is perfectly suited to
determine the molecular dynamics in the liquid phase. Usually this
dynamics is studied by splitting the spectra into diffusion and rotation contributions

\begin{equation}
\label{eqswisodiff}
S(q,\omega)=S(q,\omega)_\mathrm{trans} \otimes S(q,\omega)_\mathrm{rot}
\end{equation}

where $S(q,\omega)_\mathrm{trans}$ is the translational contribution and
$S(q,\omega)_\mathrm{rot}$ is associated with the rotation of the molecule, assuming that both movements are independent from each other.
If we assume that the translation is described by a diffusion
mechanism and, therefore, described by the Fick equation, and that
rotation is isotropic \cite{bee}:
\begin{equation}
\label{eqswdiff}
S(q,\omega)_\mathrm{trans}=\frac{1}{\pi}  \; \frac{Dq^2}{\omega^2+ \left( Dq^2
\right)^2}
\end{equation}
\begin{equation}
\label{eqswiso} S(q,\omega)_\mathrm{rot}=A_0(q\cdot
R)\delta(\omega)+\sum_{l=1}^\infty A_l(q \cdot
R)\frac{1}{\pi} \; \frac{l(l+1)D_r}{\omega^2+\left[l(l+1)D_r\right]^2}
\end{equation}
where $D$ and $D_r$ are the translational and rotational diffusion
coefficients, $A_l(q \cdot R)$ are spherical bessel functions and
$R$ is the radius of rotation. We have performed QENS experiments
at the TOFTOF spectrometer \cite{toftof} at the  FRM II reactor (Munich) in order to
determine the dynamics of 1,2-trans-dichloroethylene. The data were
corrected for self-absorption using the FRIDA software \cite{frida}. A series of fittings for each temperature with the model described by
equations \ref{eqswisodiff}, \ref{eqswdiff} and \ref{eqswiso}. Usually, each $q$ value is fitted separately, obtaining the diffusion coefficient
from a second linear fit to the $q^2$ dependence of the broadening
of the central lorentzian $ \Gamma_\mathrm{trans}(q^2)=D \cdot q^2$,
and the radius of rotation from the obtained $A_0(q\cdot
R)={sin(qR)}/{qR}$ or eventually by independently fitting all spectra
 using equation \ref{eqswisodiff} to each $S(q=q_i,\omega)$. However, our hypothesis is described by the
whole set of the aforementioned equations
\ref{eqswisodiff},\ref{eqswdiff} and \ref{eqswiso}, and thus errors arising from the two-step fitting procedure can be
minimized by simply fitting the spectra $S(q,\omega)$ for all $q$
values, i.e. fitting the complete $q$-dependent data set with only
$D$, $D_r$ and $R$ as physical parameters. The results for the radius
of rotation are shown in figure \ref{Rgirobayesg} using the presented Bayesian method together with those obtained using a Levenberg-Marquardt algorithm for each $q$-value. First of all, because fittings were performed by the frequentist approximation separately for each $q$-value (see figure \ref{Rgirobayesg}b), the radius of rotation has a $q$-dependence which is not present in the Bayesian fitting (see figure \ref{Rgirobayesg}a), consequently stressing the importance of fitting the whole data set together. A fitting  using the Bayesian algortihm has also been performed to a spectrum for $q=0.4 \mathring{A}^{-1}$ and $T=300K$ in order to compare the error bars obtained by both methods. The Error bar using the Bayesian approach was calculated by obtaining the PDF for the radius of rotation and then fitting a Gaussian function with $\sigma = \varepsilon$, being  $\varepsilon$ the frequentist parameter error. This error bar is plotted in figure \ref{Rgirobayesg}b, together with that determined by the frequentist method. As it can be seen in the figure the error obtained by the presented method is much bigger that that estimated by the frequentist method. The presented Bayesian method is therefore able to deal with simultaneous fitting of various curves, obtaining the PDF of physical parameters as a function of temperature (see figure \ref{Rgirobayesg}c,d,e).

\subsection{Model selection: Diffusion in phospholipid membranes}
Phospholipids are the main component of cell walls and can also be used in technological applications as for example drug delivery or food industry. Their dynamics is studied on many time- and length-scales with different techniques, among them quasielastic time-of-flight neutron scattering which probes the motions that dominate on times of about 100\,ps.

As will be discussed in detail elsewhere \cite{busch2009}, the question arose from previous neutron scattering experiments \cite{koenig1992,tabony1990} whether the long-range motion of phospholipids is visible on these times or if the motion appears rather localized, trapped in a cage of neighbours. This difference can be seen in the line shape of $S(q,\omega{})$: Motions that are localized during the observation time cause a central line that is not broadened beyond the resolution of the instrument but cause a foot in the spectrum. In contrast, long-range motions do broaden the central line.

\begin{figure}
\includegraphics[width=\columnwidth]{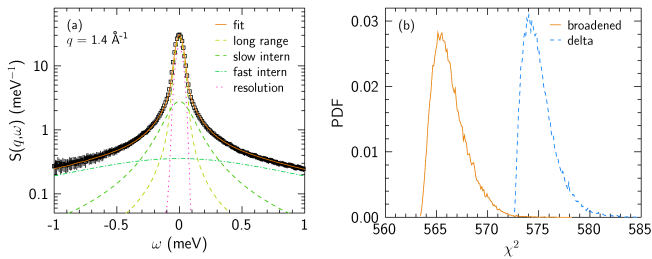}
\caption{(a) Spectra of the hydrated phospholipid DMPC together with a fit and the instrumental resolution. Internal motions were approximated with two Lorentzians, the long-range motion was either assumed to be invisible (delta-shaped central line, not shown) or visible (broadened central line, shown here). (b) The $\chi^2$ PDF associated with both the broadened central line and a delta function, showing that for any combination of parameters the broadened model is preferable compared to the delta model: The data justify the assumption that the long-range motion is visible.}
\label{membrane}
\end{figure}

The neutron scattering experiments were performed with the phospholipid DMPC (1,2-Dimyristoyl-sn-Glycero-3-Phosphocholine) in a liquid crystal fully hydrated with D$_2$O at the neutron time-of-flight spectrometer TOFTOF at the FRM II (Munich). A typical spectrum is shown in figure \ref{membrane}a after standard corrections including self absorption and subtraction of the D$_2$O spectra, obtained with the program FRIDA \cite{frida}. It is possible to fit the data "satisfactorily" with both, a broadened and a delta-shaped central line.

As stated before, Bayesian analysis is able to quantify how "satisfactory" the fits are, taking into account the whole $\chi^2\{P_i\}$ landscape and avoiding assumptions about it. In figure \ref{membrane}b, the PDFs associated with $\chi^2$ for the two models are displayed. The normal Levenberg-Marquardt algorithm would simply return the parameters at the minimal reachable value of $\chi^2$ together with this quantity. It is obvious that introducing an additional parameter, the nonzero width of the central line, reduces the $\chi^2$. The question that needs to be answered is if this reduction is significant enough to justify the additional parameter.

The $\chi_\nu^2$ gives this answer, however relying on the assumptions discussed above. Employing the Bayesian Analysis, no assumptions are made as the $\chi^2\{P_i\}$ landscape is rendered explicitly. One can see in figure \ref{membrane}b that the model incorporating a broadened central line does not only yield the smaller $\chi^2$ minimum but also the PDF associated with $\chi^2$ is for any combination of parameters smaller than the one of the delta model.

Therefore, the model comparison between the two possibilities of broadened and non-broadened central line favours the model with a broadened line.

\section{Summary}

We have proposed a general Bayesian method to  fit data, analyze results from the fit, and from these results to perform model selection between competing hypotheses. In contrast to the classical frequentist approach, where some assumptions are done concerning the $\chi^2$ landscape (there is only a minimum of $\chi^2\{P_i\}$ able to describe data within its error, this minimum has a square dependence on the parameters, and parameters are not correlated), the proposed method samples the parameter space with the
only guide of probability, thus having the following advantages:

\begin{itemize}
  \item In the fitting procedure, the Bayesian method will not get stuck
in local minima if its barrier is smaller than the error
associated with the experimental data set.
  \item Parameters are obtained as PDFs and,
because the whole parameter space is sampled, correlations between
parameters are naturally taken into account. Moreover, a natural
way to define errors based on the PDF of parameters is obtained within this method, which following the frequentist approximations would be the 68\% confidence interval around the most probable parameter value, i.e. the parameter is inside these limits with a probability P=0.68. PDFs may take an arbitrary form,
for example indicating that only a superior limit to the parameter
can be extracted from the experimental data.

  \item The likelihood (which as we have seen is directly related to $\chi^2$) obtained with this method is also a PDF hence
revealing the whole complexity of the parameter landscape. Model
selection is then performed taking into account all parameter
combinations compatible with the experiment.
    \item The presented method is flexible enough to deal with low counts
    experiments where the Poissson distribution cannot be
    approximated by a Gaussian function, by simply redefining
    $\chi^2$ in the Gibbs sampling algorithm.
\end{itemize}

This work was supported by the Spanish Ministry of Science and Technology
(FIS2008-00837) and by
the Catalonia government (2005SGR-00535).


\begin{thebibliography}{9}
\bibitem{frequentist} {The "frequentist" description defines probability of a certain event $A$ ($P(A)$) as the limiting frequency with which the event $A$ is observed when a great number of events $A$ is taken into account.}
\bibitem{diagonal}{In fact it is possible to calculate errors taking into account
the correlation between parameters using the frequentist approach (still supposing a parabolic dependence of parameters on
$\chi^2\{P_i\}$). This could be done diagonalizing the covariant
matrix. This procedure is nevertheless not usually found in the
literature resulting in an underestimation of errors. }
\bibitem{jumpchange}{Parameters are changed randomly, but their maximum change is restricted. A new parameter is therefore generated folowing $P_i^\mathrm{new}=P_i^\mathrm{old}+(\mathrm{RND}-0.5)\cdot 2P_i^{max}$ where RND is a random number between 0 and 1, and $P_i^\mathrm{max}$ is the maximum change allowed. }
\bibitem{errors}{Errors are defined supposing that parameter PDFs follow a Gaussian distribution $P(parameter)= \frac {1}{\sigma \sqrt{2\pi}}exp \left( -\frac{x^2}{2\sigma^2}\right)$. The probability that a parameter is within the interval $\pm \sigma$ is 0.683.}
\end{thebibliography}
\end{document}